# RANDOM-RESISTOR–RANDOM-TEMPERATURE KIRCHHOFF-LAW–JOHNSON-NOISE (RRRT–KLJN) KEY EXCHANGE

**Laszlo B. Kish** [1)], **Claes G. Granqvist** [2)]

[1)] *Texas A&M University, Department of Electrical and Computer Engineering, College Station, TX 77843-3128, USA*

[2)] *Department of Engineering Sciences, The Ångström Laboratory, Uppsala University, P.O. Box 534, SE-75121 Uppsala, Sweden*

**Abstract**

We introduce two new Kirchhoff-law–Johnson-noise (KLJN) secure key distribution schemes which are generalizations of the original KLJN scheme. The first of these, the Random-Resistor (RR–) KLJN scheme, uses random resistors with values chosen from a quasi-continuum set. It is well-known since the creation of the KLJN concept that such a system could work in cryptography, because Alice and Bob can calculate the unknown resistance value from measurements, but the RR–KLJN system has not been addressed in prior publications since it was considered impractical. The reason for discussing it now is the second scheme, the Random-Resistor–Random-Temperature (RRRT–) KLJN key exchange, inspired by a recent paper of Vadai, Mingesz and Gingl, wherein security was shown to be maintained at non-zero power flow. In the RRRT–KLJN secure key exchange scheme, both the resistances and their temperatures are continuum random variables. We prove that the security of the RRRT–KLJN scheme can prevail at non-zero power flow, and thus the physical law guaranteeing security is not the Second Law of Thermodynamics but the Fluctuation–Dissipation Theorem. Alice and Bob know their own resistances and temperatures and can calculate the resistance and temperature values at the other end of the communication channel from measured voltage, current and power-flow data in the wire. However, Eve cannot determine these values because, for her, there are four unknown quantities while she can set up only three equations. The RRRT–KLJN scheme has several advantages and makes all former attacks on the KLJN scheme invalid or incomplete.

Keywords: KLJN key exchange; information theoretic security; unconditional security.

## 1. Introduction

The Kirchhoff-law–Johnson-noise (KLJN) secure key distribution scheme [1–20] is a classical-statistical physical alternative to the quantum key distribution. Figure 1 depicts a binary version of the KLJN scheme and shows that, during a single-bit exchange, the communicating parties (Alice and Bob) connect their randomly chosen resistor (including its Johnson noise generator) to a wire channel. These resistors are randomly selected from the publicly known set $\{R_L, R_H\}$, $R_L \neq R_H$, representing the Low (*L*) and High (*H*) bit values. The Gaussian voltage noise generators—mimicking the Fluctuation–Dissipation Theorem and delivering band-limited white noise with publicly agreed bandwidth—produce enhanced thermal (Johnson) noise at a publicly agreed effective temperature $T_{eff}$, typically $T_{eff} >> 10^{10} \text{K}$, so that the temperature of the wire can be neglected. The noises are statistically independent of each other and from the noise of the former bit period.

In the case of secure bit exchange—*i.e.*, the *LH* or *HL* bit situations for Alice and Bob—an eavesdropper (Eve) cannot distinguish between these two situations by measuring the noise spectra $S_u(f)$, $S_i(f)$ of voltage and/or current in the cable, respectively, because the *LH* and *HL* noise levels are identical (degenerated). Thus when Alice and Bob detect the noise spectra (or noise levels) characteristic of the *LH* and *HL* situation, they know that the other party has the opposite bit and that this bit is secure. Then one of them will invert the bit (it is publicly pre-agreed who will do this) to get the same key bit as the other party. The KLJN scheme offers unconditional (information theoretic) security at both ideal and slightly non-ideal (*i.e.*, practical) conditions [3].

To avoid a potential information leak by variations in the shape of a probability distribution, the noises are Gaussian [1], and it has been proven that other distributions cannot offer perfect security [17,18]. The security against active (invasive) attacks is provided by the robustness of classical-physical quantities, which guarantees that they can be continuously monitored and exchanged between Alice and Bob via authenticated communication. Therefore the system, and the consistency of the measured and exchanged voltage and current data with the known cable parameters and model, can be checked *continuously* and *deterministically* without destroying these data, which is totally different from the case of a quantum key distribution.

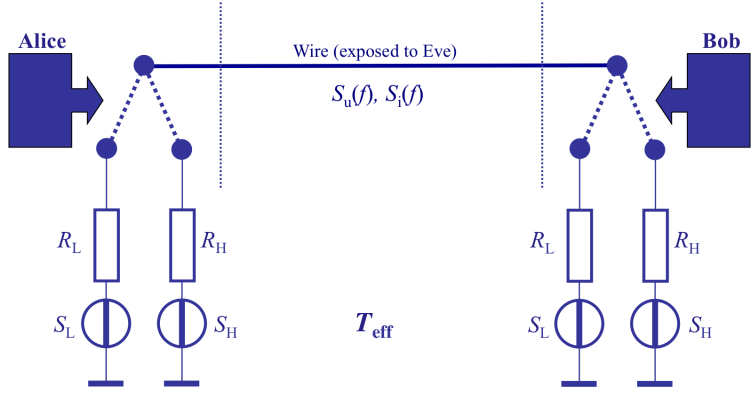

Fig. 1. Core of the KLJN scheme without defense circuitry [2] against active (invasive) attacks and attacks utilizing non-idealities. The $R_L$ and $R_H$ resistors, identical pairs at Alice and Bob, represent the Low (*L*) and High (*H*) bit values. The corresponding noise spectra $S_L$ and $S_H$ also form identical pairs at the two ends, but they belong to independent Gaussian stochastic processes. Both parties have the same temperature, and thus the net power flow is zero. The *LH* and *HL* bit situations of Alice and Bob produce identical voltage and current noise spectra, $S_u$ and $S_i$, in the wire, implying that they represent a secure bit exchange. The *LL* and *HH* bit arrangements, which occur in 50% of the cases, have singular noise levels in the wire, and hence they do not offer security because Eve can distinguish them. Consequently 50% of the bits must be discarded. This system works also with arbitrary, non-binary resistor values as an analog circuitry to exchange continuum information about the distribution of random resistors.

We must keep in mind that the KLJN secure information exchanger is basically an analog circuit and can work with arbitrary resistances because, even if the resistance values are not pre-agreed, Alice can calculate Bob's resistance from the measured data [1] by using Johnson's formula, and *vice versa*. For example, by using the measured current spectrum in the wire one obtains

$$R_B = \frac{kT_{\text{eff}}}{S_i} - R_A \quad . \tag{1}$$

It is important to note that Eve is also able to determine an arbitrary, non-pre-agreed (non-publicly known) resistor pair connected to the line by using measured voltage and current spectra [1]. The two solutions of the obtained second order equation provide two resistance values of the pair according to

$$R_{1,2} = \frac{4kTS_u \pm \sqrt{(4kTS_u)^2 - 4S_u^3 S_i}}{2S_u S_i} \quad . \tag{2}$$

However, Eve cannot determine which resistor is with Alice and which is with Bob, and hence the information exchange about the distribution of arbitrary, non-binary resistor values is secure in the original KLJN system.

## 2. The Vadai–Mingesz–Gingl KLJN scheme

Resistor inaccuracies in the binary KLJN scheme can remove the degeneracy of the *LH* and *HL* noise levels in the communication line—thus yielding non-zero information leak—as was pointed out long ago [21,22], and ensuing inaccuracies on the 1-%-level have been considered acceptable for practical purposes with minimal privacy amplification and secure bit filtering [3]. Recently, Vadai–Mingesz–Gingl (VMG) published a modified KLJN scheme [19] with very interesting properties in order to fully eliminate such a leak. We call this the VMG–KLJN system. In a subsequent article [20] they showed that the earlier temperature-compensation defense principle [16] against wire resistance attacks on the KLJN system can successfully be used also for the VMG–KLJN scheme.

We note, in passing, that the title of VMG's paper [19] is misleading because it mentions "arbitrary" resistors and indicates that the "arbitrariness" of these resistors would constitute the main new result of their paper. However, already the original KLJN scheme had "arbitrary" resistors, but not a continuum range of resistors with *ad hoc* random choice, which the VMG–KLJN scheme also is unable to offer. The truly new aspect of the VMG–KLJN scheme is different, namely that Alice and Bob can have two *different* pairs of *fixed* resistors consisting of $R_{AL}$, $R_{AH}$, $R_{BL}$, and $R_{BH}$, which are Alice's and Bob's logic *L* and *H* resistances, respectively; see Figure 2.

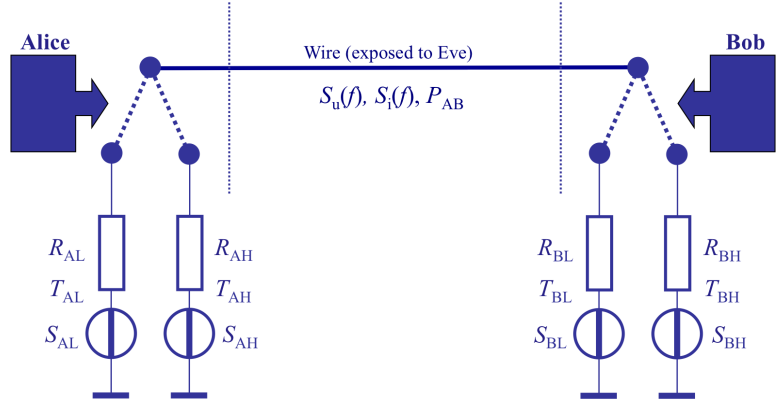

Fig. 2. Vadai–Mingesz–Gingl KLJN scheme. The resistor pairs representing the Low (*L*) and High (*H*) bit values are different at Alice (A) and Bob (B), and their temperatures and the corresponding noise spectra are different too. Thus the net power flow is non-zero. The resistance values $R_{AL}$, $R_{AH}$, $R_{BL}$ and $R_{BH}$, as well as the temperatures $T_{AL}$, $T_{AH}$, $T_{BL}$ and $T_{BH}$, are pre-determined and thus publicly known. This protocol is *not* designed to work with *arbitrary*, non-binary resistor values in order to exchange continuum information about the distribution of random resistors, because the resistance values and temperatures are interrelated, and Alice and Bob cannot abruptly alter them without requesting a change of the temperature(s) at the other party.

The VMG–KLJN scheme is binary, just as the original KLJN system, and Alice's and Bob's task is to find conditions under which the voltage and current noise spectra in the wire are identical at the *HL* and *LH* bit combinations, *i.e.*, for the pairs $R_{AL} - R_{BH}$ and $R_{AH} - R_{BL}$. This identity cannot be accomplished with uniform temperatures. Starting with the temperature $T_{AL}$ of Alice's $R_{AL}$ resistor, the other temperatures ($T_{AH}$, $T_{BL}$ and $T_{BH}$) are designed so that the *LH* and *HL* bit situations produce identical voltage and current noise spectra and power-flow in the wire, implying that they represent a secure bit exchange. VMG found the necessary temperature values [19] in the following generic form:

$$T_{AH} = T_{AL} F(R_{AL}, R_{AH}, R_{BL}, R_{BH}), \qquad (3)$$

$$T_{BL} = T_{AL} G(R_{AL}, R_{AH}, R_{BL}, R_{BH}), \qquad (4)$$

and

$$T_{BH} = T_{AL} H(R_{AL}, R_{AH}, R_{BL}, R_{BH}). \qquad (5)$$

The functions *F*, *G* and *H* are *deterministic* (their explicit forms are published [19] and are not reproduced here). Thus Alice and Bob must know not only their own set of resistors but also the resistance values at the other side, and Bob must also know Alice's temperature $T_{AL}$. Consequently these resistor sets and temperatures are deterministic, which in accordance with Kerckhoffs's principle [23] implies that all of the parameters are known by Eve. (Note that "keying" these parameter values by randomly generating and disseminating them via secure communication, by using the formerly shared key, is of course possible, just as it is the case of the Keyed-KLJN scheme [5] and some quantum versions [24,25], but such enhancements to make Eve's job more difficult are not the topic of the present paper).

The *LL* and *HH* bit arrangements, which occur in 50% of the cases, have singular noise levels in the wire. Thus they offer no security and must be discarded, so the VMG system does not offer any speed-up of the key exchange.

Concerning practical applications of the original KLJN system in "macroscopic" circuit boards or hybrid integrated circuits, the resistor pairs can easily be chosen with high-enough precision, and therefore the VMG system is not needed. However, in monolith integrated circuits and in the absence of post-processing for trimming of the KLJN resistors, the VMG method [19] can be handy to eliminate the information leak due to resistor inaccuracies [21,22].

It should be emphasized that VMG's paper [19] contains a very important discovery: that *unconditional security can be attained at non-zero power flow*, *i.e.*, at non-equilibrium conditions! Hence it is not the Second Law of Thermodynamics that guarantees security in their system but the Fluctuation–Dissipation Theorem, via the Johnson–Nyquist formula. (We note, in passing, that a similar assertion was made in the very first paper on the KLJN scheme [1]; this argument was later replaced by one involving the more widely known Second Law of Thermodynamics, which is applicable as well).

## 3. The Random-Resistor (analog) KLJN scheme

Before we turn to our main results, we outline the Random-Resistor (RR–) KLJN scheme in Figure 3. This system is not binary but analog and in thermal equilibrium. The RR–KLJN scheme employs random resistors chosen from a quasi-continuum set of resistance values. It is well-known since the inception of the KLJN concept that such a system could work, because Alice and Bob can calculate the unknown resistance value from measurements, but this system has not been addressed in prior publications as it was considered impractical.

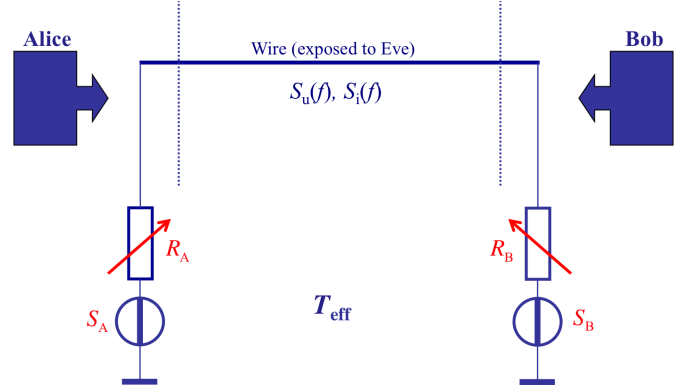

Fig. 3. Random-Resistor KLJN scheme. The temperature at the two sides is the same and is a pre-defined, publicly known constant value; thus the net power flow is zero. The resistors at Alice (A) and Bob (B), and their corresponding voltage noise spectra, are continuum random variables with a new random choice made at the beginning of each KLJN period. The Low (*L*) and High (*H*) bit values at Alice and Bob are determined by the relative resistance values; for example, the party with the higher resistance has the high bit. From voltage and current measurements, Eve can estimate the two resistance values but not their locations, unless the resistors are identical. In the hypothetical but non-practical case when the resistance distribution is a continuum, and when the inaccuracies of Alice's and Bob's estimations are zero, then 100% of secure bit exchange is accomplished because the probability of choosing two identical resistances is zero. In the practical case with finite accuracy (finite bit exchange duration) and a quasi-continuum discrete distribution, the secure bit exchange efficiency is less than 100%, because some of the bits must be discarded, but it is greater than 50%.

## 4. The Random-Resistor–Random-Temperature KLJN scheme

The important discovery by VMG [19]—that unconditional perfect security exists at non-zero power flow—is the feature that inspired our new Random-Resistor–Random-Temperature (RRRT–) KLJN; see Figure 4 which defines two new parameters, $\alpha$ and $\beta$, by $R_B = \alpha R_A$ and $T_B = \beta T_A$. Yet our new scheme is completely different from that of VMG because it uses really "arbitrary" (*ad hoc*) resistances as in the RR–KLJN scheme and moreover random (*ad hoc*) temperatures from a continuum interval. Thus the forthcoming resistance and temperature values are unknown even by Alice and Bob (analogously with their lack of knowledge about the next secure key bit), except for the ranges of values. Consequently, even Kerckhoffs's principle [23] does not allow any information leak about resistance and temperature values, and only their continuum range is publicly known. This fact makes all formerly known attack types invalid in their original form and, without further development of them, they offer zero information gain about Eve's keys.

### 3.1 Security proof of the RRRT–KLJN scheme

We analyze the protocol from Alice's point of view, which obviously is valid for Bob too.

- Known to Alice: her own temperature, resistance and noise spectrum, and the wire measurements of $S_u(f)$, $S_i(f)$ and power $P_{AB}$ flowing to Alice from Bob.
- Unknown to Alice: $\alpha$ and $\beta$.
- Known to Eve: wire measurements of $S_u(f)$, $S_i(f)$ and $P_{AB}$.
- Unknown for Eve: $\alpha$, $\beta$, $T_A$ and $R_A$.

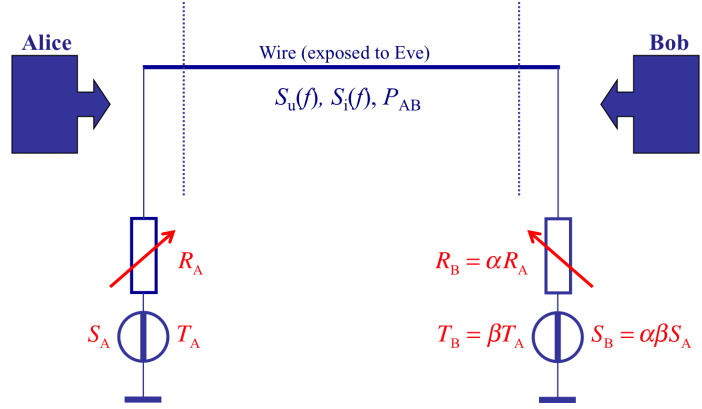

Fig. 4. Random-Resistor–Random-Temperature KLJN scheme. The temperatures and the resistors at Alice (A) and Bob (B), and their corresponding voltage noise spectra, are continuum random variables with a new random choice made at the beginning of each KLJN period. The Low ($L$) and High ($H$) bit values at Alice and Bob are determined by the relative resistance values; for example, the party with the higher resistance has the high bit. Eve cannot determine the two resistance values, not even their sum, as in the KLJN, VMG–KLJN and RR–KLJN cases. In the hypothetical but non-practical situation when the resistance distribution is a continuum and the inaccuracies of Alice's and Bob's estimation results are zero, 100% of secure bit exchange is accomplished. In the practical case with finite accuracy (finite bit exchange duration) and quasi-continuum discrete distribution, the secure bit exchange efficiency is less than 100%, because the bits with singular noise and power levels must be discarded, but 100% can be approached with proper design.

Alice wants to find out Bob's unknown parameters $\alpha$ and $\beta$. She can set up three equations by using the principles of linear operations on noise and have

$$S_{u}(f) = 4kT_A R_A \left[\frac{\alpha R_A}{R_A(1+\alpha)}\right]^2 + \alpha\beta 4kT_A R_A \left[\frac{R_A}{R_A(1+\alpha)}\right]^2 = \frac{4kT_A R_A}{(1+\alpha)^2}\alpha(\alpha+\beta) \ , \qquad (6)$$

$$S_{i}(f) = \frac{4kT_A R_A}{R_A^2(1+\alpha)^2} + \frac{4kT_A R_A \alpha\beta}{R_A^2(1+\alpha)^2} = \frac{4kT_A}{R_A}\frac{1+\alpha\beta}{(1+\alpha)^2} \ , \qquad (7)$$

and the power flow, according to earlier work [16], is

$$P_{AB} = \Delta f \frac{\alpha R_A R_A}{(R_A + R_A \alpha)^2}(\beta-1)4kT_A = 4kT_A \Delta f \frac{\alpha(\beta-1)}{(1+\alpha)^2} \ , \qquad (8)$$

where $\Delta f$ is the noise bandwidth. Equations (6)–(8) allow us to define three new quantities, which Alice can calculate from her own parameters and the measured data by

$$\gamma = \frac{S_u(f)}{4kT_A R_A} = \frac{\alpha(\alpha+\beta)}{(1+\alpha)^2} \ , \qquad (9)$$

$$\varphi = \frac{P_{AB}}{4kT_A \Delta f} = \frac{\alpha(\beta-1)}{(1+\alpha)^2} \ , \qquad (10)$$

and

$$\delta = \frac{S_\mathrm{i}(f)R_\mathrm{A}}{4kT_\mathrm{A}} = \frac{1+\alpha\beta}{(1+\alpha)^2} \ . \tag{11}$$

In can be shown that Equations (9) and (10) lead to a second-order equation for $\beta$, which has two solutions according to

$$\beta_{1,2} = \frac{-\delta(1-2\gamma)-\varphi\gamma \pm \sqrt{[\delta(1-2\gamma)+\varphi\gamma]^2 + 4(\gamma-1)(\delta-\varphi)(\varphi-\gamma\delta)}}{2(\gamma-1)} \ . \tag{12}$$

In situations for which one solution is positive and the other is negative, which is unphysical, the positive result provides Alice with Bob's temperature value. When both solutions are positive, an alternative second order equation, created from Equations (10) and (11), must be solved; the joint solution of that equation with the solution of Equation (12) yields the correct temperature parameter $\beta$ of Bob. Finally, knowing the correct value of $\beta$, any one of Equations (9)–(11) yields Bob's resistance parameter $\alpha$.

Importantly, Eve cannot determine these values because, for her, there are *four* unknown quantities while she can set up only *three* equations. Thus she has infinite possibilities, provided the continuum system is unbounded, which is impractical.

In practical applications, the solution of the above equations will not be needed, especially when we consider the fact that the RRRT–KLJN system will be a digital one, similarly to case of the KLJN realizations. This means that the temperature and resistance data will form a quasi-continuum discrete set with resolution given by the bit resolution of the system. Thus instead of the calculations outline above, a bottom-up version is feasible and practical as a consequence of its reduced calculation need during operation. This approach involves a tabulation of all possible temperature and resistor settings at Alice and Bob and creating a look-up table from the data on $S_\mathrm{u}(f)$, $S_\mathrm{i}(f)$ and $P_\mathrm{AB}$.

One should note that this kind of tabulation must be done in any case in order to locate possible singular combinations of $S_\mathrm{u}(f)$, $S_\mathrm{i}(f)$ and $P_\mathrm{AB}$ that could uniquely inform Eve about the resistance and temperature situations. For any secure bit, the measurable set of $S_\mathrm{u}(f)$, $S_\mathrm{i}(f)$ and $P_\mathrm{AB}$ must be degenerated and thus must occur for at least two opposite bit situations within the statistical inaccuracy of the KLJN operation; otherwise Eve can extract the bit by using her own model of the system, which she can build according to the Kerckhoffs's principle [23]. Those singular shared bits are insecure and must be discarded during operation whenever they occur.

*3.2 Immunity against former attacks*

The RRRT–KLJN scheme has several advantages and makes all existing and previously valid attacks invalid in their known form. For example, the key exchange speed is virtually doubled because, with proper design, almost all bit exchange period supply a secure key bit due to overlap of the noise and power levels belonging to different bit settings. Resistor- or temperature-inaccuracies do not matter, and they can no longer be utilized for Hao-type attacks [22,26]. The Bergou–Scheuer–Yariv–Kish cable resistance attack [27,28] is also invalid in its known form, just as is the new cable-capacitance attack by Chen *et al*. [29]. Finally, the new transient attack [30] by Gunn–Allison–Abbott does not work either because of the unknown resistances and temperatures.

New attacks are of course possible. For example, if Eve compares the mean-square voltages at the two ends of the wire she gets a new equation, and then she probably has enough equations to extract information due to the non-zero wire resistance. An information leak will then exist, and the real question is: how large is this leak considering Eve's poor statistics due to the strongly limited bit-exchange period; see related analysis elsewhere [3,7].

### *3.3 Some practical considerations*

A disadvantage of the RRRT–KLJN scheme is that the Kish–Granqvist temperature-compensation defense mechanism [16] cannot be used to nullify cable resistance effects against an as yet unknown attack type of such kind. A perhaps more practical version of the RRRT–KLJN scheme is the generalization of the formerly proposed Multiple–KLJN (MKLJN) system [5] based on a random choice of a known large set of resistors by introducing a random choice of temperatures from a known large set of such data. As already mentioned above, the known sets must be properly checked, because only choices with degenerated voltage/current/power values can be considered secure–not the singular values. Bit-error analysis [13,14] and error removal is still an open problem in the RRTT–KLJN scheme.

**Conclusions**

We introduced two schemes with arbitrary (*ad hoc*) random resistor choices and enhanced communication speed. The RRRT–KLJN scheme also has *ad hoc* random temperatures, and this makes the new scheme unique among the existing KLJN versions because even the sum of the resistances is secret. All former attacks are invalid and, as a minimum, need further developments to extract any information.

The RRRT system also has disadvantages, and some advanced features of the enhanced KLJN schemes, such as the iKLJN [5], cannot be used and some of the defense features against active (invasive) attacks may need to be upgraded.

Only the future can tell if the RRRT scheme remains a topic of purely academic interest or whether it will lead to important practical applications. Both the generation of a random (analog) resistance and a random temperature look technically feasible, especially since accuracy and reproducibility of the resistance values are unimportant. There is no reason to use the RR–KLJN scheme because the RRRT scheme needs only a minor expansion: controlling the mean-square amplitude of the noise voltage generators.

It is yet an open question if the original KLJN system and its enhanced versions iKLJN, KKLJN, VMG, *etc.*, or the new RRRT–KLJN scheme, is more feasible for practical applications; all of them offer unconditional (information theoretic) security.

**Acknowledgements**

Discussions with Zoltan Gingl are appreciated.